\documentclass[runningheads]{llncs}
\usepackage{graphicx}
\usepackage{hyperref}
\usepackage{cleveref}
\usepackage{array}
\usepackage[shortlabels]{enumitem}
\usepackage{caption}
\usepackage{subcaption}
\usepackage{multirow}
\usepackage{xcolor,colortbl}
\usepackage{pgfplots} % package used to implement the plot  
\usepackage{comment}
\usepackage{tabularx}
\usepackage{rotating}
\usepackage{adjustbox}

\pgfplotsset{width=6.5cm, compat=1.6,
    /pgfplots/ybar legend/.style={
    /pgfplots/legend image code/.code={%
       \draw[##1,/tikz/.cd,yshift=-0.25em]
        (0cm,0cm) rectangle (3pt,0.8em);},
   },
}
\usetikzlibrary{patterns}

\definecolor{Gray}{gray}{0.85}
\definecolor{LightCyan}{rgb}{0.88,1,1}

\newcolumntype{a}{>{\columncolor{Gray}}c}
\newcolumntype{b}{>{\columncolor{white}}c}

\newcommand{\etal}{\textit{et al}. }
\newcommand{\ie}{\textit{i}.\textit{e}. }
\newcommand{\eg}{\textit{e}.\textit{g}. }

\usepackage{fancyhdr}
\fancypagestyle{firststyle}
{
\fancyhead{}

\fancyfoot[CE,CO]{\scriptsize M. Rebol et al., "CPR Emergency Assistance Through Mixed Reality Communication"\\The Version of Record of this contribution is published in Augmented Intelligence and Intelligent Tutoring Systems and is available online at \href{https://doi.org/10.1007/978-3-031-32883-1_38}{https://doi.org/10.1007/978-3-031-32883-1\_38}.
}
\fancyfoot[LE,RO]{\thepage}
}

\begin{document}
\title{CPR Emergency Assistance Through Mixed Reality Communication}
%
%\titlerunning{Abbreviated paper title}
% If the paper title is too long for the running head, you can set
% an abbreviated paper title here
%
\author{Manuel Rebol\inst{1,2,3}\orcidID{0000-0003-3846-4356} \and
Alexander Steinmaurer\inst{1}\orcidID{0000-0002-1760-2855} \and
Florian Gamillscheg\inst{1}\orcidID{0000-0003-0470-6644} \and
Krzysztof Pietroszek\inst{2}\orcidID{0000-0002-5127-9801} \and
Christian Gütl\inst{1}\orcidID{0000-0001-9589-1966} \and
Claudia Ranniger\inst{3}\orcidID{0000-0002-2630-020X} \and
Colton Hood\inst{3}\orcidID{0000-0002-4249-4294} \and
Adam Rutenberg\inst{3}\orcidID{0000-0002-6471-1050} \and
Neal Sikka\inst{3}\orcidID{0000--0002-4696-6922}
}

\authorrunning{M. Rebol et al.}
% First names are abbreviated in the running head.
% If there are more than two authors, 'et al.' is used.
%
% \institute{Princeton University, Princeton NJ 08544, USA \and
% Springer Heidelberg, Tiergartenstr. 17, 69121 Heidelberg, Germany
% \email{lncs@springer.com}\\
% \url{http://www.springer.com/gp/computer-science/lncs} \and
% ABC Institute, Rupert-Karls-University Heidelberg, Heidelberg, Germany\\
% \email{\{abc,lncs\}@uni-heidelberg.de}}
\institute{Graz University of Technology, Austria \and
American University, USA \and
George Washington University, USA \\
 \email{rebol@gwu.edu}\\
 %\url{https://mrebol.github.io/} %\and
 %ABC Institute, Rupert-Karls-University Heidelberg, Heidelberg, Germany\\
 %\email{\{abc,lncs\}@uni-heidelberg.de}
 }

\maketitle              % typeset the header of the contribution
\thispagestyle{firststyle}
\begin{abstract}
We design and evaluate a mixed reality real-time communication system for remote assistance during CPR emergencies. Our system allows an expert to guide a first responder, remotely, on how to give first aid. RGBD cameras capture a volumetric view of the local scene including the patient, the first responder, and the environment. The volumetric capture is augmented onto the remote expert's view to spatially guide the first responder using visual and verbal instructions.
We evaluate the mixed reality communication system in a research study in which participants face a simulated emergency. The first responder moves the patient to the recovery position and performs chest compressions as well as mouth-to-mask ventilation. 
Our study compares mixed reality against videoconferencing-based assistance using CPR performance measures, cognitive workload surveys, and semi-structured interviews.
We find that more visual communication including gestures and objects is used by the remote expert when assisting in mixed reality compared to videoconferencing. Moreover, the performance and the workload of the first responder during simulation do not differ significantly between the two technologies.

\keywords{Mixed Reality \and CPR \and Remote collaboration.}
\end{abstract}
%
%
%% A teaser figure can be included as follows
\begin{figure}
     \centering
     \begin{subfigure}[b]{0.38\textwidth} % use this number to set the space between the pictures
         \centering
         \includegraphics[height=2.9cm]{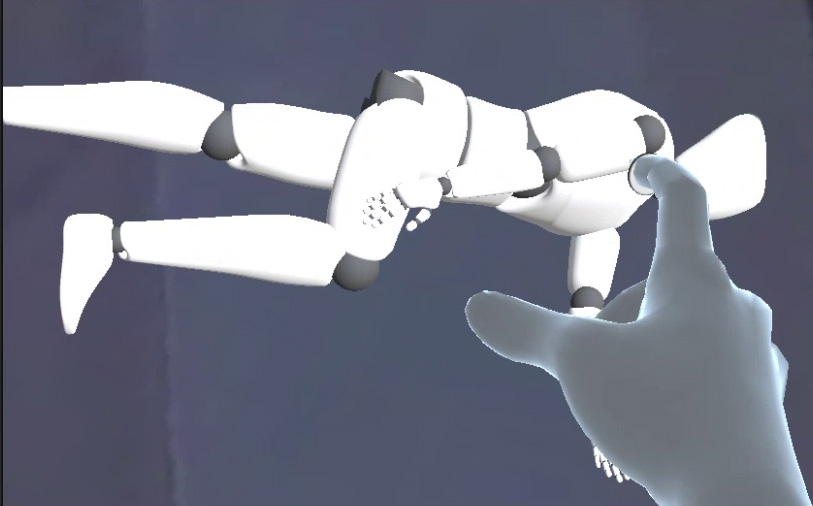}
         \caption{Pointing and 3D model}
         \label{fig:pointing}
     \end{subfigure}
     %\hfill
     \begin{subfigure}[b]{0.19\textwidth}
         \centering
         \includegraphics[height=2.9cm]{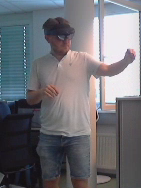}
         \caption{Instructor}
         \label{fig:instructor}
     \end{subfigure}
     %\hfill
     \begin{subfigure}[b]{0.39\textwidth}
         \centering
         \includegraphics[height=2.9cm]{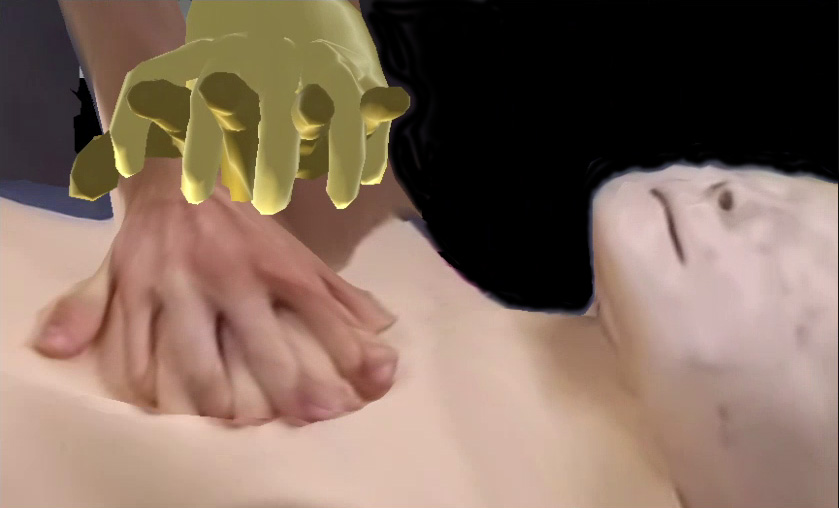}
         \caption{Hand technique and position}
         \label{fig:hands}
     \end{subfigure}
        \caption{First aid assistance through mixed reality. The instructor (b) points at a 3D holographic model that demonstrates the recovery position (a) in the augmented field of view of the first responder. The instructor uses a 3D holographic hand model (c) for guidance when giving chest compressions to the patient.}
        \label{fig:teaser}
\end{figure}
\section{Introduction} 
In recent years interest and popularity in Augmented Reality (AR) devices have increased drastically and have attracted the attention of the research world and consumers alike \cite{Carmigniani:2010:ARTSA}. AR has many possible applications that range from entertainment and education to design and manufacturing \cite{Carmigniani:2010:ARTSA}, \cite{Kamphuis:2014:ARIME}, \cite{Nee:2012:ARAIDAM}. AR can be described as a technique that augments the real world with virtual computer-generated information or content with which the user is able to interact \cite{Azuma:2001:RAIAR} \cite{Carmigniani:2010:ARTSA}. As opposed to Virtual Reality (VR) approaches, where users are fully immersed in the virtual environment and have no connection to the real world, AR tries to superimpose virtual objects upon the user's surroundings \cite{Carmigniani:2010:ARTSA}. AR technology is not restricted to head-mounted displays (HMD) but is also used in handheld devices like smartphones and tablets \cite{Azuma:2001:RAIAR}. The virtual objects that are embedded in the real world provide information to the user, which can help in performing tasks, such as by helping workers through the electrical wire system of an aircraft \cite{Carmigniani:2010:ARTSA}, \cite{Mizell:2000:ARAIA}. In the past few years AR has progressed immensely and more and more AR applications and devices have entered the market. 
% AR in medicine
In medicine, AR technology has enormous potential, where HMDs are for example an essential tool in computer-aided surgery (CAS) \cite{Birkfellner:2002:AHMOBFAR}. In this field, AR can provide useful information to surgeons in a less distracting manner, compared to information on a monitor \cite{Birkfellner:2002:AHMOBFAR}. In addition to surgical settings, AR has been used in endoscopic procedures \cite{Mahmud:2015:CVARIGE}, needle puncture procedures \cite{Kim:2012:EVPPIA} and in training for complex surgical procedures \cite{Coles:2011:IHARFPNI}. Aside from practical applications, AR has gained a lot of attention in the field of medical education like anatomy, where teaching methods have not changed significantly over the last century. \cite{Duarte:2020:LABVRAR}. In the medical field, learning predominantly involves workplace training, which is time-consuming, not very cost-efficient, and comes with some risk \cite{Kamphuis:2014:ARIME}. AR can offer a safer learning environment, where concepts can be practiced without the fear of making errors \cite{Kamphuis:2014:ARIME}. One area where proper training is particularly important is cardiopulmonary resuscitation (CPR) \cite{Cave:2011:IITCRAD}. Cardiac arrest is a significant public health problem, with approximately 350,000 people per year in the US and Canada alone receiving CPR \cite{Travers:2010:P4COAHAG}. Well-executed and timely CPR has a significant impact on survival and neurological outcome, which is why in 2011 the American Heart Association (AHA) published a recommendation for mandatory CPR training starting at school-age \cite{Cheng:2015:PCQICF}, \cite{Cave:2011:IITCRAD}. However, multiple studies have shown that even when carried out by healthcare professionals, CPR quality is often poor \cite{Abella:2005:QCRDIHCA}, \cite{Gallagher:19995:EOBCRS}. With its visual feedback, AR technology can be a valuable tool for guiding responders. 

In this research project, we introduce a mixed reality system, where the user is guided by a remote expert in performing first aid (\Cref{fig:teaser}). Our RGBD cameras allow for three-dimensional visual information that helps the user through the steps. Within this project, we conducted an evaluation with 30 participants separated into two groups. Both groups were given the same tasks. They had to give first aid to a lifeless person, bring the person into the recovery position, and start with CPR after the person stopped breathing. We compared instruction via the mixed reality (MR) approach (group A) with video-based communication (group B). We analyzed objective metrics of CPR quality recorded by the CPR mannequin and data from users including workload surveys and interviews. We conclude our main contributions as follows: 

\begin{enumerate}
    \item We introduce an MR communication system designed for remote first aid assistance.
    \item We conducted a comparison between MR communication technology and video-based communication.
    \item The project team measured workload and performance when giving assisted first aid in MR and videoconferencing.
\end{enumerate}
 
The paper is structured as follows: \Cref{sec:rel-work} gives an overview of past research on Extended Reality (XR) applications for first aid assistance. In \Cref{sec:mr-design}, we discuss the design of the MR communication system, focusing on the different views and their interaction with each other and the devices and software that were used. We evaluate the presented MR communication system in a research study described in \Cref{sec:research-study}. After introducing the study procedure, we present and discuss the results that were obtained. In the end, we summarize the work and give an outlook on future work.

\section{Related Work} \label{sec:rel-work}

%\subsection{Extended Reality Communication}
Extended Reality (XR) \cite{rebol-xr} has been a valuable communication tool %in domains like construction \cite{Broschart:2013:ARCTUDP} and 
especially in medicine. In medical education, AR-supported instructions assist students in demanding tasks by offering a realistic learning environment in which they can develop their theoretical knowledge through didactics, as well as improve their practical skills through interactive simulation \cite{Tang:2019:ARIME}, \cite{Christopoulos:2022:TEARSI}. In the field of telemedicine, a field that uses communication technology to diagnose and treat patients remotely, AR has developed great potential. Wang \etal \cite{Wang:2017:ARTMPRPT} developed a telemedicine mentoring application using Microsoft's HoloLens. Their application allows remote experts to guide trainees through complex medical situations with hand gestures that were then displayed in the AR environment of the trainee. Their study, which examined the usability in a trauma setting revealed that their AR setup was regarded as more immersive than other telemedicine approaches. Similarly, Lin \etal \cite{Lin:2018:AFMSPAR} used AR in a surgical setting in combination with a Microsoft HoloLens. Their study, which used a lower-leg fasciotomy as a training task, revealed that participants who used the headset received a higher performance score and reported higher usability. In surgical settings, AR allows surgeons to better concentrate on the operating field in comparison to traditional telemedical approaches, where surgeons are often forced to shift their focus to screens for receiving assistance from an expert, which can result in more errors \cite{Andersen:2017:ARBAST}. Andersen \etal \cite{Andersen:2017:ARBAST} used a different approach, which features a tablet PC that is located between the local surgeon and the patient. The tablet captures live video that is sent to the remote expert. The remote expert can then annotate the video with graphical information, which is then sent back to the local surgeon's screen. The user study found that participants who used the AR approach completed the tasks with higher accuracy and fewer distractions. 
Besides surgical settings, XR technology has been also used for consultation purposes. Anton \etal \cite{Anton:2017:ATPFRMC} used a combination of AR and VR devices to build a telemedicine system for remote consultations. Their setup consists of an AR client, that captures 3D surface information of the environment, which is then transmitted to the remote expert via a communication module, which enables peer-to-peer connection. On the physician side, the VR client receives the streamed data and is responsible for the rendering on a 3D display, which allows the physician to examine the information and interact with it. In the field of postoperative care, Ponce \etal \cite{Ponce:2016:TWMDARREPC} used AR on mobile devices in their study to let physicians virtually examine the patient over a long distance. Their application allows users to interact with each other via mobile devices \eg with visual annotations on the patient's screen. Their user study revealed that the 96\% of the patients regarded the setup as useful, while physicians were slightly less satisfied with 89.6\% of them expressing that the application was useful.

%Talk about different AR communication systems used in medicine and for other applications. 

\subsection{Extended Reality for First Aid Assistance}

Multiple studies have demonstrated the potential of extended reality for assisting users in first aid.
Already in 1997 Zajtchuk and Satava \cite{Zajtchuk:1997:MAVR} pointed out the potential of VR for medical education. Kuyt \etal \cite{Kuyt:2021:TUVRARECPR} included 42 articles in their review on the evolution of XR approaches for CPR training. Their study indicated that the number of both VR and AR applications for resuscitation training is rapidly growing. Due to the high rise of AR-related CPR publications and advancements in technology in recent years, the 
proportion of AR-related articles is expected to rise in the future. The study concludes that XR shows great potential for CPR training environments, which will likely result in innovations and novel applications.
As an example, Girau1 \etal \cite{Girau:2019:AMRSSFA} proposed a VR-based training application for first aid, using an HTC Vive headset and a real mannequin to provide haptic feedback. Similarly, Blome \etal \cite{Blome:2017:VRNGLVR} also described a VR setup that uses a non-verbal approach for teaching and practicing reanimation. In contrast to VR, where users are fully immersed in their virtual environment, AR shifts the focus of the user's interaction to the real world \cite{Barsom:2016:SREARAMT}. It follows that AR applications can represent tasks in the real world more realistically than their VR-based counterparts. Some studies have already shown the potential of AR in medical education and first aid assistance. Fromm \etal \cite{Fromm:2019:TPARIOFA} developed a concept for an AR application that teaches users first aid. The study revealed that an AR application would help users in emergency situations in an intuitive and quick manner. Johnson \etal \cite{Johnson:2018:HCPRDEMRI} used Microsoft's HoloLens for building \textit{HoloCPR}, an AR application that provides real-time instructions for CPR. The subsequent evaluation revealed how the use of such devices can result in a better reaction time and improved accuracy. Frøland \etal \cite{Frøland:2020:SOAFDUAR} also used Microsoft HoloLens for developing a training environment for trauma first aid. In their study, they simulated an emergency, where a patient suffered from severe bleeding that had to be stopped by the participants.

%%%%%%%%%%%%%%%%%%%%%%%%%%%%%%%%%%%%%%%%%%%%%%%%%%%%%%%%%%%%%%%%%%%%%%%%%%%%%%%%%%%%%%%%%%%%%%%%%%%%%%%%%%%%%%%%%%%%%%%%%%%%%%%%%%%%%%
\section{Designing Mixed Reality Communication for First Aid Assistance} \label{sec:mr-design}
We designed the mixed reality (MR) communication system in consultation with a first aid instructor. The domain expert described the information required to understand the emergency situation. Moreover, the domain expert suggested how visual communication can be supported using virtual objects. The detailed system design approach and implementation details approach were described in \cite{rebol-hicss} and \cite{rebol-ismar}, respectively.

\begin{figure*}[h]
     \centering
         \centering
         \includegraphics[width=\textwidth]{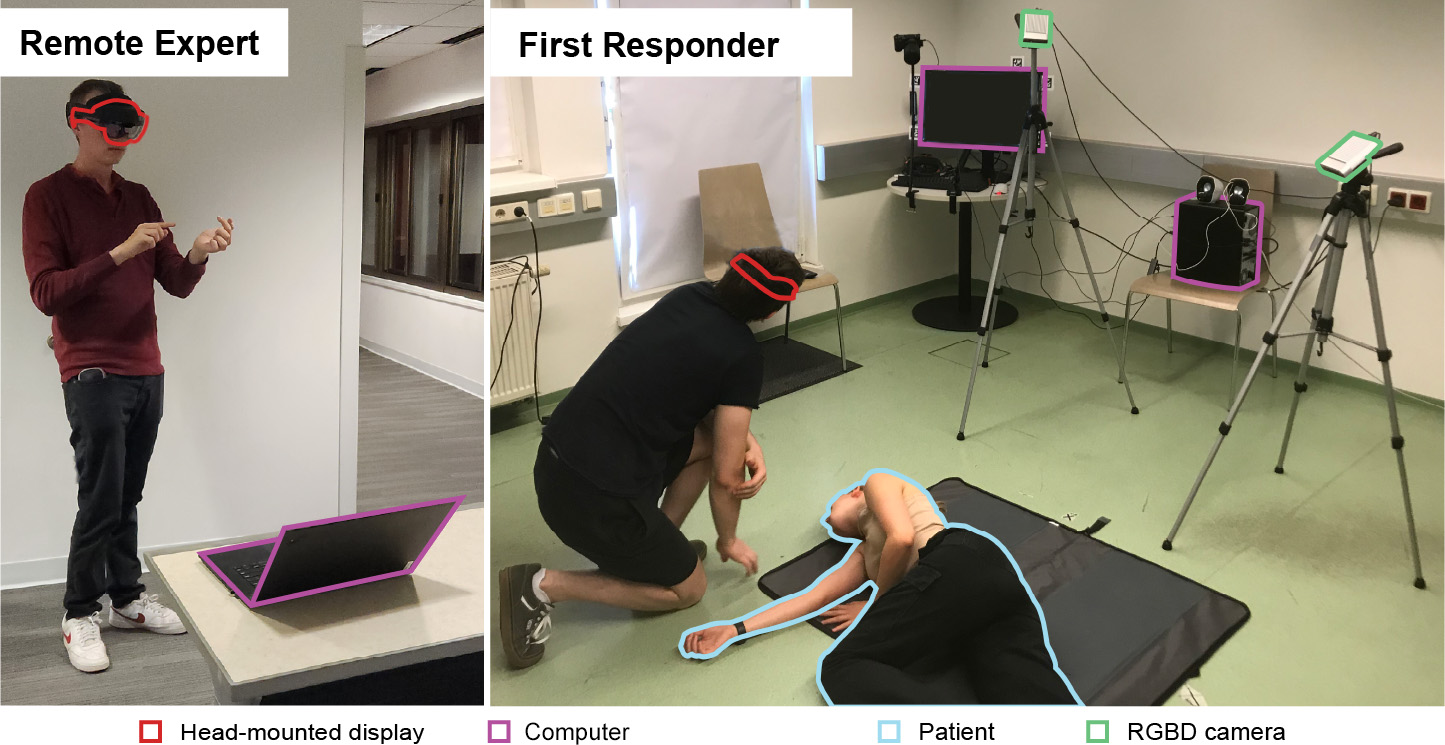}
         \caption{MR system components. %We outline the devices used for the mixed reality communication system. 
         The first responder's scene (right) is captured using RGBD cameras (green) and sent over to the remote expert (left). The remote expert uses the 3D information to guide the first responder using head-mounted displays (red). Computers (purple) handle the network communication and rendering.}
         \label{fig:mr-system}
\end{figure*}

\subsection{Augmented Views}
Both the expert and the first responder wear the Microsoft Hololens 2 head-mounted display (HMD) to communicate. In addition, to the view, the HMD offers head and hand tracking. Head tracking is needed to anchor augmented objects in the physical environment. Thus, alignment between the views and visual interaction on a shared view is possible. Hand tracking allows for gesture communication between the users of the system.

\paragraph{Remote Expert}
The remote expert's view is dominated by a mesh view of the local first responder's scene. The view is captured by Microsoft Azure Kinect RGBD cameras at the local scene. The remote expert can switch between two different RGBD cameras to get a volumetric view of the local scene. The RGBD cameras are positioned such that they capture the patient, the first responder, and the environment. In future versions of the system, the HMD-included camera can be used to capture the local scene instead of the separate RGBD camera. We used a separate Azure Kinect camera because the current Microsoft Hololens depth sensor does not provide sufficient quality.   

In addition to the mesh view, the remote expert is presented with an augmented video feed from the local scene. The $1920 \times 1080$ video feed provides the remote expert with a high-quality view of the local scene including details the mesh view misses because of limitations of the time-of-flight technology. 

Besides the views of the local scene, the remote expert sees augmented objects which they can use to visually guide the first responder. The remote expert can manipulate the objects, \ie resize, rotate, move, using gestures. 

\paragraph{First Responder}
The first responder's view is dominated by the physical environment. The augmented information for the first responder is kept to a minimum such that they can focus on the emergency. The augmented information for the first responder consists of augmented objects, the expert's augmented hands, and a video feed showing the remote expert. The first responder's MR communication role is passive. They see augmented objects and screens but do not actively manipulate them. 

\subsection{Interaction}
The remote expert and the first responder can verbally communicate using audio. Moreover, the mixed reality system supports visual communication using gestures and case-specific augmented objects. For first aid, we provide a 3D holographic model that demonstrates the rescue position, hands illustrating the chest compression position, and an object to show the depth of the chest compressions. Only the remote expert actively manipulates augmented objects and screens in MR using gestures. The first responder manipulates the physical environment. We show the devices used given the research study setup in \Cref{fig:mr-system}.

\section{Research Study} \label{sec:research-study}
We evaluated the proposed MR system in a research study in which we compare it against videoconferencing-based communication. Our research focused on understanding how MR communication can be used in medical emergencies. We measured the workload between video-only and MR assistance and analyzed how it differs. Moreover, we were interested in whether higher performance can be achieved when using MR for emergency assistance compared to video-only communication in the context of first aid / CPR.

\begin{figure}
     \centering
         \centering
         \includegraphics[width=\columnwidth]{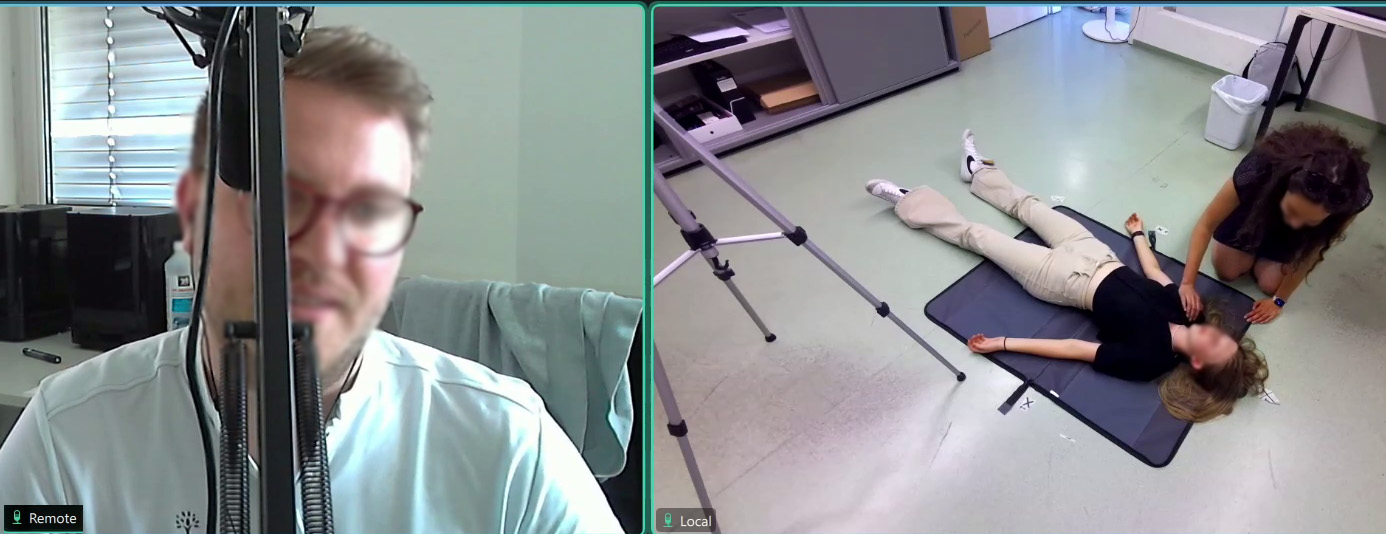}
         \caption{CPR assistance through videoconferencing. A remote expert (left) guides the first responder (right) through the emergency using %state-of-the-art 
         video communication. }
         \label{fig:video-conferencing}
\end{figure}

\subsection{Study Design} % The overall design and the instruments

\paragraph{Subjects} Altogether, 30 participants were recruited as first responders. The study participants did not receive information about the content of the training prior to arrival at the test site, to prevent preparation and more closely simulate an unexpected emergency. Informed consent was obtained, and subjects were randomized to either test condition.   All training sessions were conducted by a single expert who has over ten years of professional experience in first-aid training.

\paragraph{Room Setup} Expert and first responder were placed in separate rooms to ensure that all communications occurred through the specified modality; room setup remained otherwise static for both conditions. Each room contained a patient actor who simulated an unconscious but still breathing patient, a CPR mannequin (Laerdal Little Anne QCPR mannequin) capable of measuring metrics of CPR quality that are aligned with AHA guidelines \cite{little-anne-qcpr}, and a mask for ventilation. We show the MR setup in \Cref{fig:mr-system}, and videoconference setup (Cisco Webex) in \Cref{fig:video-conferencing}. 

\begin{figure}
\newcommand{\myheight}{2.5cm}
    \centering
    	\captionsetup[subfigure]{position=bottom, font={stretch=0}}
         \subfloat[\small{Recovery position}]{\includegraphics[height=\myheight]{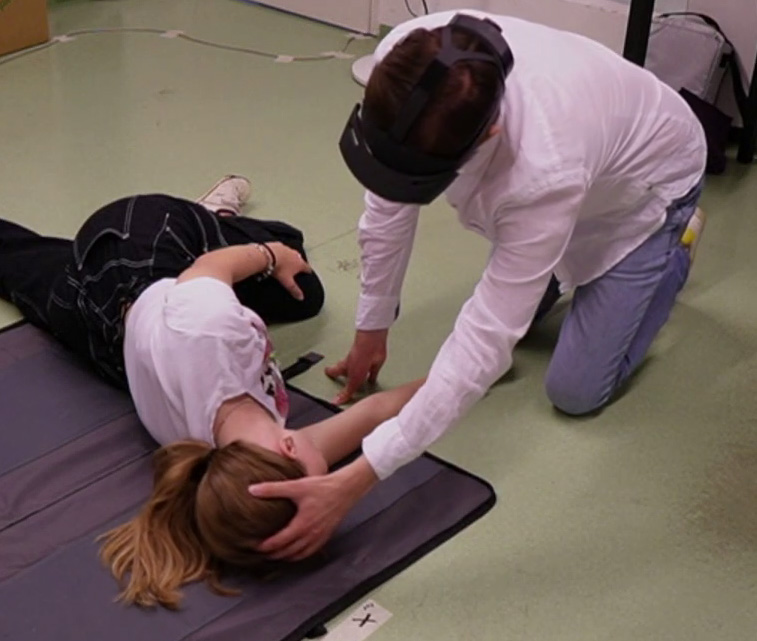} }
         \subfloat[\small{Breathing check}]{\includegraphics[height=\myheight]{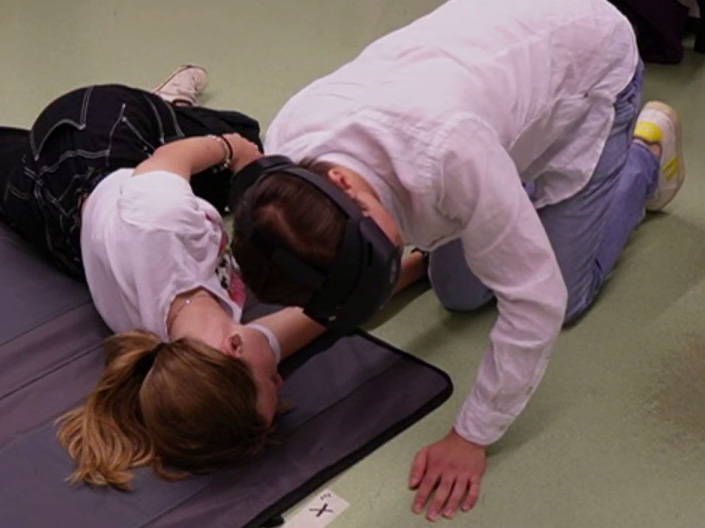} }
         \subfloat[\small{Chest compres.}]{\includegraphics[height=\myheight]{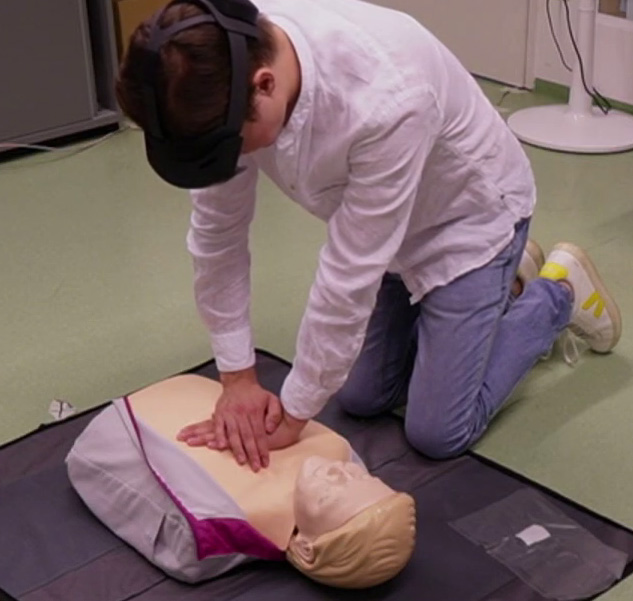} }
         \subfloat[\small{Ventilation}]{\includegraphics[height=\myheight]{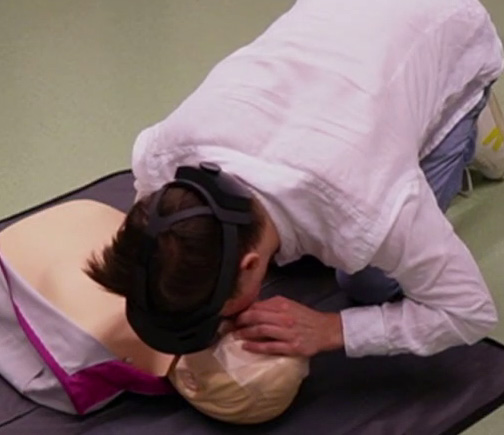} }
        \caption{First aid steps. We present the 4 main steps of the first aid simulation. The first responder performs steps (a) and (b) on a patient actor and steps (c) and (d) on a CPR mannequin.}
        \label{fig:first-aid-steps}
\end{figure}

\paragraph{First Aid Training}  We illustrate the four main steps actively performed by the first responder in \Cref{fig:first-aid-steps}. The responder first assessed the patient actor to determine that they were unconscious but breathing, and placed the patient in the recovery position. The responder then continued to monitor the patient's breathing. After a defined period the patient actor held their breath to simulate cardiac arrest, at which time a CPR mannequin replaced the patient. The responder then provided chest compressions and mouth-to-mask ventilation. In total each participant was instructed to give CPR for 4 minutes.   Metrics of CPR quality such as rate and depth of compression were recorded from the mannequin during this time.

\paragraph{Surveys} At the conclusion of the simulation, participants completed a demographic survey, the NASA Task Load Index (NASA-TLX) \cite{NASATLX:1988} and the Simulation Task Load Index (SIM-TLX) \cite{SIMTLX:2020}, both metrics of cognitive workload. Finally, the MR group also completed an MR-specific questionnaire.  The expert instructor completed NASA- and SIM-TLX once after completing the training.

\subsection{Results and Discussion}

\paragraph{Subject Demographics} We illustrate the first responder demographics in \Cref{tab:demographics}. 

\begin{table}
	\setlength{\tabcolsep}{12pt}
	%\scriptsize
	%\footnotesize
 	\caption{First responder demographics and prior experience.
	}%}
	\centering
	\begin{tabular}{l||c|c}
		%\textbf{Attribute} & \\ \hline \hline
        {} & \textbf{Video} & \textbf{AR} \\ 
		\hline \hline
		{Age} & 23(7) y & 23(4) y \\ \hline
		{Male/Female} & 11/4 & 11/4\\ \hline
		{No. of prior first aid exposures} & 5(9)& 4(6)\\ \hline
		{No. of prior MR experiences} & 0& 0\\ \hline \hline
	\end{tabular}
 \label{tab:demographics}
\end{table}	

\paragraph{CPR Performance}
 Data for 25 learners (13 video and 12 AR) was included; 5 data samples were removed due to mannequin malfunction. 
There was no significant difference in CPR performance when comparing video and MR conditions for any CPR performance category (2 sample, 2 tailed t test with significance of $\alpha=0.05$). The data is presented in \Cref{tab:cpr-data}. We noticed a trend towards better ventilation and worst compression score in AR.

\begin{table}
	\setlength{\tabcolsep}{0.175cm}
     \caption{CPR mannequin data. Mean ($\bar{x}$) and standard deviation ($\sigma$) for mannequin-derived CPR performance metrics, in Video and AR.  %We present the mannequin evaluation metrics (second column) in four categories (first column). In the third and fourth columns, we compare the technologies of video and mixed reality (MR), respectively, for first aid. %The mean, standard deviation of the two groups are shown. The better mean result is highlighted in bold. 
    The last column compares the two groups by presenting the p-value of a Student's t-test.}
    %}
    \centering
    \begin{tabular}{l|l||a|c||a|c||c}
        ~ & ~ & \multicolumn{2}{c||}{\textbf{Video}} & \multicolumn{2}{c||}{\textbf{AR}} & \textbf{T-test} \\ 
        \textbf{Category} & \textbf{Measurement [unit]} & \textbf{$\bar{x}$}  & \textbf{$\sigma$} &  \textbf{$\bar{x}$} & \textbf{$\sigma$}  & {p-val.} \\ \hline \hline
        \multirow{2}{*}{\textbf{Overall}} & Score [\%] & 59\% & 20\% & 56\% & 23\% & 0.76 \\ \cline{2-7}
        ~ & Flow Fraction [\%] & 60\% & 6\%  & 60\% & 10\%  & 0.97 \\ \hline
        \multirow{2}{*}{\textbf{Transition}} & Heart & 33 & 5 & 34 & 8 & 0.74 \\ \cline{2-7}
        ~ & Lung & 1.0 & 0.7  & 1.3 & 0.9 & 0.31 \\ \hline
        \multirow{6}{*}{\textbf{Compress.}} & Compression Score [\%] & 82\% & 17\% & 68\% & 19\% & 0.07 \\ \cline{2-7}
        ~ & Number of Compressions [n] & 253 & 33 & 266 & 40 & 0.41 \\ \cline{2-7}
        ~ & Avg (n/min) & 109 & 17 &  114 & 18 & 0.48 \\ \cline{2-7}
        ~ & Correct Release [\%] & 92\% & 17\% & 80\% & 28\% & 0.22 \\ \cline{2-7}
        ~ & Correct Depth Rate [\%] & 96\% & 8\%  & 94\% & 12\%  & 0.66 \\ \cline{2-7}
        ~ & Compr. w/ adeq. speed [\%] & 42\% & 36\%  & 44\% & 40\%  & 0.87 \\ \hline
        \multirow{4}{*}{\textbf{Ventilation}} & Ventilation Score [\%] & 48\% & 31\% & 55\% & 25\% & 0.53 \\ \cline{2-7}
        ~ & Number of ventilations [n] & 7 & 4 & 10 & 6 & 0.23 \\ \cline{2-7}
        ~ & Adequate ventiliation [\%] & 99\% & 5\% & 99\% & 3\% & 0.99 \\ \cline{2-7}
          & Hyperventilation [\%]  & 1\%  & 5\%  & 1\%  & 3\%  & 0.99 \\ \hline \hline
    \end{tabular}
    \label{tab:cpr-data}
\end{table}

\paragraph{Workload}
We illustrate the overall results of both workload surveys, NASA-TLX and SIM-TLX, for first responders and the expert in \Cref{tab:results}. The expert reported a lower overall workload for both surveys when using the MR technology compared to videoconferencing. The first responders reported similar workload between the two technologies. A two-sample two-tailed t-test was not able to show a significant difference ($\alpha = 0.05$) between the groups. The mean overall workload is similar when comparing the two technologies.

% \begin{table}[]
% \centering   
% \setlength\extrarowheight{1pt} 
% 	\setlength{\tabcolsep}{18pt}
% \begin{tabular}{l|c|c} 
% \hline \hline \bf{Cognitive Load} & \bf{Mean} & \bf{SD} \\ \hline
% Mental Demand & 86 &  6\\
% Physical Demand & 50 & 29  \\
% Temporal Demand & 22 & 24\\
% Performance & 66 & 22 \\
% Effort & 79 & 11 \\
% Frustration & 47 & 31 \\
% \hline \hline
% \end{tabular}
% \caption{\label{tab:nasa} NASA-TLX Scores.}
% \end{table}

% \begin{table}[]
% \centering
% \setlength\extrarowheight{1pt} 
% 	\setlength{\tabcolsep}{18pt}
% \begin{tabular}{l|c|c} 
% \hline \hline \bf Cognitive Load & \bf{Mean} & \bf{SD} \\ \hline
% Mental Demand & 72 & 13\\
% Physical Demand & 52 & 24  \\
% Temporal Demand & 21 & 20\\
% Frustration & 48 & 30 \\
% Task Complexity & 72 & 17\\
% Situational Stress & 40 & 28\\
% Distractions & 16 & 16\\
% Perceptual Strain & 22 & 17\\
% Task Control & 58 & 23\\
% \hline \hline
% \end{tabular}
% \caption{\label{tab:sim} SIM-TLX Scores.}
% \end{table}

\definecolor{blue1}{RGB}{100,180,255}
\definecolor{red1}{RGB}{255, 60, 60}
\definecolor{green1}{RGB}{0, 193, 0}
\begin{figure}[!ht]
%\vspace{-0.2cm}
        \centering
        %\resizebox{\columnwidth}{!}{%
%\begin{adjustbox}{trim=0 0 0 1cm}%
\raisebox{0.25cm}{\begin{tikzpicture}
%\vspace{-4cm}
%\edef\mylst{"A","B","C","D","E","F","G","H","i","j","k","l","m","n", "o","p","q","r","s","t","u","v"}
\begin{axis}  
[  bar width=4, % NEW BIT
    ybar, % ybar command displays the graph in horizontal form, while the xbar command displays the graph in vertical form.  
    enlargelimits=0.12,% these limits are used to shrink or expand the graph. The lesser the limit, the higher the graph will expand or grow. The greater the limit, the more graph will shrink.   
    legend style={at={(0.6,1.11)}, % these are the measures of the bottom row, where -0.25 is the gap between the bottom row and the graph.   
      anchor=north, legend columns=2},     
      % here, north is the position of the bottom legend row. You can specify the east, west, or south direction to shift the location.   
    ylabel={NASA-TLX score}, % there should be no line gap between the rows here. Otherwise, latex will show an error.  
    ylabel style={},
    yticklabel style={rotate=90},
    ymax=19,
    symbolic x coords={Mental D.,Physical D., Temporal D., Performance D., Frustration, Effort},  
    xtick=data, 
    xtick pos=bottom,
    xticklabel style={rotate=90},
    %nodes near coords=\pgfmathsetmacro{\mystring}{{\mylst}[\coordindex]}\mystring,  
    nodes near coords align={horizontal}, 
    nodes near coords style={font=\small, rotate=90},
    %width=0.45\columnwidth,
    height=6cm,
    ]  
% video
\addplot[fill={green1}] coordinates {(Mental D., 10) (Physical D., 17) (Temporal D., 8) (Performance D., 7) (Frustration, 3) (Effort, 6)}; % these are the measures of a particular bar graph. The tick marks of the y-axis will be adjusted automatically according to the data values entered in the coordinates.  
\node [above,xshift=0.07cm,yshift=0.4cm, style={font=\tiny, rotate=90}] at (axis cs:  Mental D., 10) {10 (9)};
\node [above,xshift=0.07cm,yshift=0.4cm, style={font=\tiny, rotate=90}] at (axis cs:  Physical D., 17) {17 (8)};
\node [above,xshift=0.07cm,yshift=0.4cm, style={font=\tiny, rotate=90}] at (axis cs:  Temporal D., 8) {8 (7)};
\node [above,xshift=0.07cm,yshift=0.4cm, style={font=\tiny, rotate=90}] at (axis cs:  Performance D., 7) {7 (6)};
\node [above,xshift=0.07cm,yshift=0.4cm, style={font=\tiny, rotate=90}] at (axis cs:  Frustration, 3) {3 (5)};
\node [above,xshift=0.07cm,yshift=0.4cm, style={font=\tiny, rotate=90}] at (axis cs:  Effort, 6) {6 (7)};

% MR
\addplot[postaction={pattern=north east lines}, fill={green1}] coordinates {(Mental D., 12) (Physical D., 14) (Temporal D., 10) (Performance D., 5) (Frustration, 2) (Effort, 7)}; 
\node [above,xshift=0.31cm,yshift=0.4cm, style={font=\tiny, rotate=90}] at (axis cs:  Mental D., 12) {12 (11)};
\node [above,xshift=0.31cm,yshift=0.4cm, style={font=\tiny, rotate=90}] at (axis cs:  Physical D., 14) {14 (9)};
\node [above,xshift=0.31cm,yshift=0.4cm, style={font=\tiny, rotate=90}] at (axis cs:  Temporal D., 10) {10 (7)};
\node [above,xshift=0.31cm,yshift=0.4cm, style={font=\tiny, rotate=90}] at (axis cs:  Performance D., 5) {5 (4)};
\node [above,xshift=0.31cm,yshift=0.4cm, style={font=\tiny, rotate=90}] at (axis cs:  Frustration, 2) {2 (4)};
\node [above,xshift=0.31cm,yshift=0.4cm, style={font=\tiny, rotate=90}] at (axis cs:  Effort, 7) {7 (7)};
\legend{Video, Mixed Reality}  
\end{axis}  
 \end{tikzpicture}  }
 %\end{adjustbox}
%       \caption{NASA TLX in-person. We compare the mean weighted NASA TLX score for learner (solid) and instructor (line pattern) teaching the US-CVC procedure. The score is shown on the y-axis and the workload categories on the x-axis. On top of the bars, mean and variance are presented.
%       }
%       \label{fig:person-nasa-tlx}
%\end{figure}
%
%\begin{figure}
%    \centering
%    %\resizebox{\columnwidth}{!}{%
\begin{tikzpicture}  
%\vspace{1cm}
\begin{axis}  
[  bar width=4, % NEW BIT
    ybar, % ybar command displays the graph in horizontal form, while the xbar command displays the graph in vertical form.  
    enlargelimits=0.08,% these limits are used to shrink or expand the graph. The lesser the limit, the higher the graph will expand or grow. The greater the limit, the more graph will shrink.   
    legend style={at={(0.6,1.1)}, % these are the measures of the bottom row, where -0.25 is the gap between the bottom row and the graph.   
      anchor=north, legend columns=2},     
      % here, north is the position of the bottom legend row. You can specify the east, west, or south direction to shift the location.   
    ylabel={SIM-TLX score}, % there should be no line gap between the rows here. Otherwise, latex will show an error.  
    ylabel style={},
    yticklabel style={rotate=90},
    ymax=12.1,
    yshift=8cm,
    symbolic x coords={Mental D.,Physical D., Temporal D., Frustration, Task Complexity, Situational Stress, Distractions, Perceptual Strain, Task Control},  
    xtick=data, 
    xtick pos=bottom,
    xticklabel style={rotate=90},
    %nodes near coords,  
    nodes near coords align={horizontal}, 
    nodes near coords style={font=\tiny, rotate=90},
    %width=0.6\columnwidth,
    height=6cm,
    ]  
% video
\addplot[fill={green1}] coordinates {(Mental D., 6) (Physical D., 11) (Temporal D., 6) (Frustration, 4) (Task Complexity, 2) (Situational Stress, 6) (Distractions, 2) (Perceptual Strain, 1) (Task Control, 2)}; % these are the measures of a particular bar graph. The tick marks of the y-axis will be adjusted automatically according to the data values entered in the coordinates.  
\node [above,xshift=0.07cm,yshift=0.35cm, style={font=\tiny, rotate=90}] at (axis cs:  Mental D., 6) {6 (5)};
\node [above,xshift=0.07cm,yshift=0.35cm, style={font=\tiny, rotate=90}] at (axis cs:  Physical D., 11) {11 (6)};
\node [above,xshift=0.07cm,yshift=0.35cm, style={font=\tiny, rotate=90}] at (axis cs:  Temporal D., 6) {6 (5)};
\node [above,xshift=0.07cm,yshift=0.35cm, style={font=\tiny, rotate=90}] at (axis cs:  Frustration, 4) {4 (4)};
\node [above,xshift=0.07cm,yshift=0.35cm, style={font=\tiny, rotate=90}] at (axis cs:  Task Complexity, 2) {2 (2)};
\node [above,xshift=0.07cm,yshift=0.35cm, style={font=\tiny, rotate=90}] at (axis cs:  Situational Stress, 6) {6 (5)};
\node [above,xshift=0.07cm,yshift=0.35cm, style={font=\tiny, rotate=90}] at (axis cs:  Distractions, 2) {2 (2)};
\node [above,xshift=0.07cm,yshift=0.35cm, style={font=\tiny, rotate=90}] at (axis cs:  Perceptual Strain, 1) {1 (1)};
\node [above,xshift=0.07cm,yshift=0.35cm, style={font=\tiny, rotate=90}] at (axis cs:  Task Control, 2) {2 (1)};

% AR
\addplot[postaction={pattern=north east lines}, fill={green1}] coordinates {(Mental D., 6) (Physical D., 9) (Temporal D., 5) (Frustration, 2) (Task Complexity, 2) (Situational Stress, 6) (Distractions, 4) (Perceptual Strain, 4) (Task Control, 3)}; 
\node [above,xshift=0.31cm,yshift=0.35cm, style={font=\tiny, rotate=90}] at (axis cs:  Mental D., 6) {6 (7)};
\node [above,xshift=0.31cm,yshift=0.35cm, style={font=\tiny, rotate=90}] at (axis cs:  Physical D., 9) {9 (6)};
\node [above,xshift=0.31cm,yshift=0.35cm, style={font=\tiny, rotate=90}] at (axis cs:  Temporal D., 5) {5 (6)};
\node [above,xshift=0.31cm,yshift=0.35cm, style={font=\tiny, rotate=90}] at (axis cs:  Frustration, 2) {2 (4)};
\node [above,xshift=0.31cm,yshift=0.35cm, style={font=\tiny, rotate=90}] at (axis cs:  Task Complexity, 2) {2 (3)};
\node [above,xshift=0.31cm,yshift=0.35cm, style={font=\tiny, rotate=90}] at (axis cs:  Situational Stress, 6) {6 (5)};
\node [above,xshift=0.31cm,yshift=0.35cm, style={font=\tiny, rotate=90}] at (axis cs:  Distractions, 4) {4 (5)};
\node [above,xshift=0.31cm,yshift=0.35cm, style={font=\tiny, rotate=90}] at (axis cs:  Perceptual Strain, 4) {4 (6)};
\node [above,xshift=0.31cm,yshift=0.35cm, style={font=\tiny, rotate=90}] at (axis cs:  Task Control, 3) {3 (2)};
\legend{Video, Mixed Reality}  
%\legend{In-person Learner, In-person Instructor}  
  
\end{axis}  
\end{tikzpicture} 

       \caption{First responder workload per category. We compare the NASA-TLX score (top) and the SIM-TLX score (bottom) presented as mean and (standard deviation) of the first responder when getting assistance through video (solid) and mixed reality (line pattern). %The score is shown on the y-axis and the workload categories on the x-axis. On top of the bars, mean and variance are presented.
       }
       \label{fig:responder-tlx}
\end{figure}
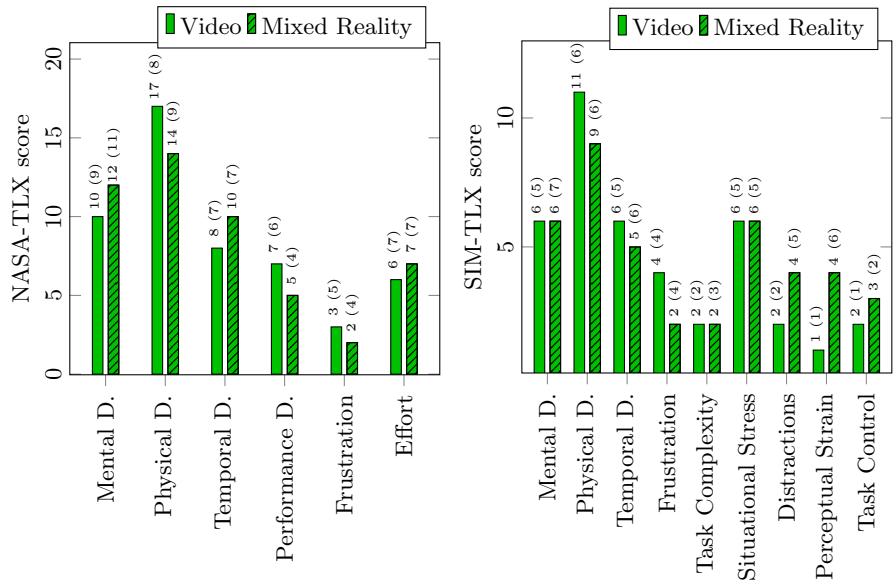

We present the per-category workload results for first responders in \Cref{fig:responder-tlx} to give a more in-depth insight. The physical demand was the highest during the first aid emergency. We posit that this is related to the physical exertion during chest compressions. 
%Overall, the mean workload reported was similar between the categories. Statistical significance cannot be shown given an adequate significance level of $\alpha = 0.05$. 
The SIM-TLX signals three interesting trends. 
% higher frustration
The video-assisted first responders reported higher frustration but lower distraction and perceptual strain than the MR communications group.% using MR communication. 
We propose that the higher frustration with video results from two factors.  First, instructions involving physical space must be communicated via 2D video and voice, which must then be translated into actions, rather than by using virtual objects to provide visual demonstration of the necessary action. Secondly, the first responders must switch gaze from the task at hand to the video, in order to compare their own progress with the instructor's directions.  %
%difficulty conveying complex instructions over video. It is harder to convey instructions visually using 2D video. It requires more effort and multiple attempts by the expert until the local responder understands the instructions. The first responder does not only have to understand the complex instruction but also has to learn from looking at a video screen compared to in-person or MR visual communication which can be frustrating. 
In AR, visual communication is more similar to in-person because the views are aligned, and pointing gestures and augmented object demonstrations can be used.    
% higher distractions and perceptual strain
%The MR- group first responders reported higher distractions and perceptual strain than the video group. We argue that
The higher perceptual strain reported by the MR group may also be due to the relative increase in visual information presented to and processed by the learner in the MR mode. Higher distraction ratings in the MR group may result from the limited field of view (FOV) of the Hololens 2 headset, and the need to locate or track instructor input outside the FOV.
%and the fact that the first responders saw them appear and disappear from their FOV. 
We argue that this can be improved by only showing them the remote hands when necessary, for example when the remote expert decides to point or gesture actively.

\begin{table}[!ht]
	\setlength\extrarowheight{2pt} 
	\setlength{\tabcolsep}{7pt}
	%\scriptsize
	%\footnotesize
    \caption{NASA-TLX and SIM-TLX workload presented as mean and (standard deviation) for first responders and the expert during first aid emergency simulations using videoconferencing and MR technology. %We highlight the lower workload means in bold.
	}
 %}
	\centering
	\begin{tabular}{r||c|c||c|c}
		%\hline \hline
		&\multicolumn{2}{c||}{\textbf{NASA-TLX}} & \multicolumn{2}{c}{\textbf{SIM-TLX}}  \\ 
		&Video & MR & Video & MR\\ \hline \hline
		%\textbf{First responder} &$ 51 \pm 12 $& $\mathbf{49 \pm 14}$ &      \textbf{$\mathbf{39 \pm 15}$} & $42 \pm 17$\\ \hline
        \textbf{First responder} &$ 51 (12) $& ${49 (14)}$ &      \textbf{${39 (15)}$} & $42 (17)$\\ \hline
		\textbf{Expert} & $49$ & \textbf{${35}$} &            $39$ & \textbf{${30}$}\\ \hline \hline
	\end{tabular}
	%\vspace{-6pt}% TODO: just to fit the 8pages
 \label{tab:results}
\end{table}	

\paragraph{Mixed Reality Survey}
%open comments from questionnaire.
The first responders noted in the open-ended questions that they especially liked the gestural communication including pointing by the remote expert. They highlighted the importance of visual communication in stressful situations. When asked about problems with the MR system, some participants noted that the visual instructions were not always given at the best position such that they had to look around to see the augmented instructions.
% remote expert
When asked about what they liked about the experience, the expert reported: ``I really liked the possibility to have additional holograms illustrating hand positions or rescue positions. This helped to provide faster help and spend less time with instructions. Additionally, the different camera perspectives were great for evaluating the quality of the CPR (chest compressions and ventilation).'' The negative aspects were: ``The initial setup of the 2D and 3D areas was sometimes a bit cumbersome. When moving the cube (a handle for moving the location of the visual feeds) the area did not move adequately. Sometimes the connection got lost due to some reason but restarting the system was not a big deal.'' The expert was also asked about features that could be added: ``Some participants complained that audio and visual instructions can be a bit overwhelming. Therefore, a visual way to show the frequency of chest compressions would be great. Especially for the rescue position, a 3D object with movable joints (knee, arm, etc.) would be helpful to show how to bring a person to the rescue position (instead of an object of the already correct position).''

\paragraph{Discussion} Although the application of MR devices seems to be futuristic within the context of emergency situations, it enables interesting perspectives. Many countries have successfully implemented guided first aid instructions over an emergency call. A (3D) video channel could provide additional information to the emergency call center regarding the set measures and their quality. 

Although the MR and video groups did not differ significantly in terms of CPR performance in this study, the use of MR with truly novice first responders may be more beneficial during complex maneuvers, for instance, to evaluate and demonstrate head tilt-chin lift for ventilation.
%The differences between the two groups, to be more precise between the video- and MR-based system is not significant when it comes to the CPR data from the mannequin. When comparing specific features such as ventilation the numbers are better (even if not statistically significant), since the additional 3D information can be helpful to see if the head is tilted correctly and the chest is rising adequately. 

Similar to AEDs, which are usually close to an emergency situation in urban areas, MR devices could be co-located with them to initiate an emergency call or request additional support from the emergency call center. 

Analyzing the combined performance and workload results, we conclude that the technology is well-suited for emergencies because the usage is intuitive. The remote expert only required brief guidance on how the MR technology works. Furthermore, the first responders put on their HMD only about 30 seconds before the actual simulation started. %This shows us that the technology is well-suited for emergencies. 
Moreover, assuming that the technology has a learning curve, better MR results can be expected after using the technology longer or more frequently.

\paragraph{Broader Impact} MR technologies are used for communication across industries. Since first aid emergencies are stressful and complex procedures, we believe that the technology can not only be used for other medical procedures including medical training, but also across domains. An example could be remote repair emergencies where a local operator needs to receive guidance from a remote expert to complete the task as quickly and accurately as possible.
Similar to giving CPR, spatial information and visual communication including pointing, gestures, and augmented objects seem invaluable.

\subsection{Limitation}
%No headset in real-world. Still relevant for training. Remote training. 
One of the main concerns of this research study is its feasibility for deployment in real-life emergency situations. However, due to the rapid development and spread of AR technologies such devices and application areas are increasing. We believe that as HMDs become lighter and slimmer, sooner than later people will wear MR devices in their everyday lives. Thus, what now seems futuristic, will become relevant in the near future. %Our MR approach should also present a showcase for (remote) training. 

%Setup takes some time which is problematic in emergency.
The current setup requires some time and experience to be installed. In emergency situations this is not possible. However, the setup can be faster and require fewer devices as technology progresses. For example, instead of Azure Kinect RGBD cameras, the Hololens 2 integrated RGBD cameras could be used to transmit spatial information to the remote expert. This would make the system more portable and easy to deploy in emergencies. Similarly, communication computers can be replaced by handheld phones or better HMD-integrated hardware. 

%Only on local network tested, although designed for truly remote.
The current research study was conducted on a local network. However, the MR system was designed for remote data transfer over the internet. Especially, given recent 5G developments, bandwidth and latency deficiencies have already been resolved in many parts of the world. 

% Natural light interference in volumetric view, ToF limitation. -> Technology, ML, recorded.
The limitation of low 3D mesh quality when the system is used outdoors because of the interference of natural infrared light with the time of flight sensor technology can be tackled by using prerecorded volumetric information. Moreover, machine learning can be used to fill in missing information and provide the remote expert with a higher-quality 3D mesh view.

\section{Conclusion}
We presented the design and the evaluation of a mixed reality (MR) communication system for first aid. The system allows a remote expert to guide a local first responder through giving first aid. Compared to help over phone and videoconferencing, the MR system allows for augmented visual instructions such as gestures, annotations, and object demonstrations. Moreover, the remote instructor is presented with a volumetric view which gives them spatial information, important for various medical procedures.  

We evaluated the MR system in a research study in which we compared MR against videoconferencing. We found that overall, the results including objective CPR mannequin performance and subjective workload measures remain similar between the technologies. We identified many new opportunities that MR offers for an expert to visually guide a first responder. The results show that visual guides do help first responders.   

In future work, we will analyze how MR can be used for other medical procedures. Moreover, we are interested in combining prerecorded procedural guidance including avatars (\cite{rebol-vr}, \cite{rebol-chi}) for the local operator with active help from the remote expert in problematic situations. This would standardize the process and allow the remote expert to save energy by reducing repetitive instructing and only intervening in critical situations when the first responder needs additional guidance. \\

%% if specified like this the section will be committed in review mode
%TODO:
%\section*{Acknowledgements}
\noindent {\footnotesize
\textbf{Acknowledgements.} The work is supported by National Science Foundation grant no. 2026505 and
2026568.}
%
% ---- Bibliography ----
%
% BibTeX users should specify bibliography style 'splncs04'.
% References will then be sorted and formatted in the correct style.
%
%\bibliographystyle{splncs04}
\bibliographystyle{splncs03} % Refs save space
\bibliography{main}
\end{document}